# Three-dimensional printing of silica-glass structures with submicrometric features

Miku Laakso[1,*], Po-Han Huang[1,*], Pierre Edinger[1], Oliver Hartwig[2], Georg S. Duesberg[2], Carlos Errando-Herranz[1], Göran Stemme[1], Kristinn B. Gylfason[1], and Frank Niklaus[1]

[1] KTH Royal Institute of Technology, School of Electrical Engineering and Computer Science, Division of Micro and Nanosystems, Sweden
[2] Institute of Physics, EIT 2, Universität der Bundeswehr Munich, Germany
[*] These authors contributed equally to this work

**Humanity's interest in manufacturing silica-glass objects extends back over three thousand years[1]. Silica glass is resistant to heating and exposure to many chemicals, and it is transparent in a wide wavelength range. Due to these qualities, silica glass is used for a variety of applications that shape our modern life, such as optical fibers in medicine[2] and telecommunications[3]. However, its chemical stability and brittleness impede the structuring of silica glass, especially on the small scale. Techniques for three-dimensional (3D) printing of silica glass, such as stereolithography[4,5] and direct ink writing[6], have recently been demonstrated, but the achievable minimum feature size is several tens of micrometers[7]. While submicrometric silica-glass structures have many interesting applications, for example in micro-optics[8], they are currently manufactured using lithography techniques, which severely limits the 3D shapes that can be realized. Here, we show 3D printing of optically transparent silica-glass structures with submicrometric features. We achieve this by cross-linking hydrogen silsesquioxane to silica glass using nonlinear absorption of laser light followed by the dissolution of the unexposed material. We print a functional microtoroid resonator with out-of-plane fiber couplers to demonstrate the new possibilities for designing and building silica-glass microdevices in 3D.**

Three-dimensional (3D) micro- and nanostructures have many exciting applications in optics, mechanics, fluidics, biology, and medicine; and they can be versatilely created using sub-picosecond laser pulses[9]. For producing 3D structures out of silica glass, a subtractive approach relies on a combination of laser exposure and chemical etching of silica-glass substrate[10–12]. Not unlike sculpting, this approach allows structures only inside the original substrate volume. Additive 3D printing is routine for polymer structures using two-photon polymerization. The printed polymer can be coated with silica glass using conformal deposition, and this approach has been used for manufacturing periodic photonic crystals[13]. However, this approach entails severe design limitations because the printed polymer structures define only the empty volumes inside the deposited silica glass, and the removal of the polymer requires access to all the polymer structures from the outside of the grown silica glass. Similarly to conformal deposition, it is possible to mold a silica nanocomposite material around 3D-printed polymer structures[14]. However, the nanocomposite contains significant amounts of carbon, which must be removed and the remaining silica glass densified to obtain a transparent object. This requires sintering at 1300 °C, resulting in severe shrinkage, which is destructive for objects attached to a substrate. Two-photon polymerization is not limited to purely organic polymers. Using hybrid organic-inorganic polymers, the two-photon polymerization acting on the organic component allows the production of structures consisting of metals[15], ceramics[16], glass-ceramics[17–19], and materials where the inorganic component consist of only silicon and oxygen[20–22]. As with nanocomposites[4,14], the removal or reduction of the organic content via post-exposure baking results in severe shrinkage, which in turn distorts structures attached to a substrate[17,18,21].

Here, we show that sub-picosecond laser pulses can selectively cross-link hydrogen silsesquioxane (HSQ), with the empirical formula[23] of $HSiO_{1.5}$, to silica glass within the HSQ bulk. Notably, HSQ is completely inorganic, and its cross-linking is not based on two-photon polymerization of an organic component. The selective cross-linking within the HSQ is possible due to the sub-picosecond pulse duration, which allows nonlinear absorption in the laser focus without simultaneous linear absorption in the HSQ outside the focal volume[9]. Cross-linking of HSQ based on linear absorption has earlier been demonstrated using an electron beam[24] and deep UV light[25]. However, because of the linear absorption, these methods essentially can only produce 2D structures or inverted 2.5D structures based on controlling the depth of linear absorption[24]. In contrast, we use a laser wavelength of 1040 nm, which HSQ does not linearly absorb.

Our glass 3D-printing process starts by drop-casting HSQ dissolved in organic solvents on a fused silica substrate (Fig. 1(a)). Once the solvents have evaporated, the dry HSQ is patterned through the transparent substrate by tracing out the desired 3D shape using sub-picosecond laser pulses (Fig. 1(b)). After laser patterning, the unexposed HSQ is removed in a development step using an aqueous potassium hydroxide (KOH) solution, leaving behind the patterned 3D structures (Fig. 1(c)).

To demonstrate the capabilities of our 3D-printing method, we fabricated different types of 3D structures (Fig. 1(d-f)). The minimum lateral width of a printed structure was approximately 0.5 µm while the minimum height was approximately 3 µm. This difference is attributed to the shape of the laser focus, which is extended in the laser propagation direction and extends in length with increased single-pulse energy[19,20]. We were able to reduce the height of the printed features to the submicrometric level by reducing the single-pulse energy, but the achieved quality of the printed structures was not as good as with higher single-pulse energy (Extended data fig. 1).

We used Raman spectroscopy to determine how closely the printed material resembles pure silica glass. Raman spectra were collected from as-printed material and from the printed material after baking it at various temperatures. These spectra were compared to a spectrum collected from a commercial silica-glass substrate. The Raman spectrum of the as-printed material contained the main spectral features of commercial silica glass, but it also contained some additional features, originating from water, hydroxyl groups, and carbon species (Fig. 2(a)). However, all the differences in Raman spectra disappeared after baking the printed HSQ structures at 900 °C or above.

Since HSQ does not contain carbon, we assume that the measured carbon species originated from the organic solvents in the HSQ solution, which might not completely have evaporated before laser patterning. The carbon concentration of a separate as-printed sample was measured to be below one percent using energy-dispersive X-ray spectroscopy (EDS). According to the Raman spectra, the carbon was removed during baking between 500 °C (Extended data fig. 2) and 900 °C. This is consistent with earlier reports on carbon-containing silicon oxides, which show a decrease in mass when baked in air at 500 °C[18,20,21]. This mass loss has been linked to the decrease of carbon-related vibrational spectroscopy signal[20].

In addition to the presence of carbon, there are other differences in the Raman spectra, in comparison to pure silica glass, for samples baked at temperatures below 900 °C. One of these spectral differences is the presence of a Si-H bond, which indicates incomplete cross-linking of HSQ. The Si-H signal disappeared already after baking at 150 °C (Extended data fig. 2). Baking an HSQ film in an oxygen atmosphere has been reported to gradually remove Si-H bonds through oxidation at temperatures between 340 °C and 650 °C[26,27]. However, the relative concentration of oxygen in comparison to hydrogen increases with decreasing film thickness[26], and for thin films with a thickness of 120 nm, the Si-H signal has been reported to disappear completely after baking at 150 °C[28]. This may indicate that the Si-H signal in the 3D-printed structures originates from a thin surface layer incompletely cross-linked during laser patterning. The Raman spectra also contain signs of molecular water and hydroxyl groups (OH), which are often observed in silica glasses with high water content[29,30]. The signals from the molecular water and hydroxyl groups significantly decrease after baking at 300 °C. The hydroxyl groups are not visible in the Raman spectra after baking at 900 °C, which is consistent with an earlier reported strong decrease in the hydroxyl-group signal after baking silica-glass precursor at 800 °C[31].

A photoluminescent background in the Raman spectra is observed in samples baked at temperatures from 150 °C up to 800 °C and significantly reduces the visibility of the Raman peaks for samples baked at 500 °C and 800 °C (Extended data fig. 2). Separate photoluminescence characterization of the sample baked at 500 °C revealed a broad photoluminescence peak slightly above 2 eV with a tail at higher energies (Extended data fig. 3). The observed photoluminescence can originate from at least three different defect types, all reported earlier for laser-damaged silica glass[32–34]. These defect peaks are 1.9 eV and 2.0 eV from non-bridging oxygen hole centers without and with hydrogen bonding[35,36], 2.2 eV from silicon clusters[37], and around 2.7 eV from oxygen-deficiency centers meaning a direct silicon-silicon bond in a silica-glass network[36]. Similarly to photoluminescence, the increase of the intensity of the Raman peaks associated with 3- and 4-membered rings in the silica-glass network have also been reported to occur after exposing the silica glass to laser[33]. These changes in the Raman spectra were reported to disappear after baking at 900 °C for one hour, which correlated with the disappearance of visible modifications in the silica glass created by the laser pulses[34]. This is consistent with the results of the present study where reordering to the standard silica-glass network occurs during baking at 900 °C.

We investigated the porosity, homogeneity, and crystallinity of the printed silica glass structures using transmission electron microscopy (TEM). We did not observe any pores in the scanning TEM images and the material was largely homogenous (Fig. 2(b)). This was corroborated by scanning electron microscope (SEM) images of a cross-section of a printed structure (Extended data fig. 4). An electron diffraction pattern confirmed that the printed material is amorphous (Fig. 2(c)). High-resolution TEM (HRTEM) images revealed a low concentration of inhomogeneities with the size of a few nanometers (Fig 2(d)). These inhomogeneities appeared crystalline (Extended data fig. 5), but due to their small size, we were unsuccessful in determining their elemental composition using EDS and electron energy loss spectroscopy (EELS). However, the compositional contrast in the HRTEM images indicates that the nanocrystals consist of heavier elements than the silica glass. The nanocrystals could be metal contamination from gold and platinum layers deposited on the TEM sample during sample preparation.

As baking at 900 °C is required for producing pure silica-glass structures, the shrinkage of the printed structures at this temperature is of interest because strong shrinkage distorts structures attached to a substrate at multiple points[17,18,21]. We characterized the shrinkage by using five 3D-printed T-shaped structures (Extended data Fig. 6). The lengths of the horizontal beams of the T-shaped structures were measured before baking and after baking at different temperatures. The mean values of the relative linear shrinkages of the five structures are shown in Fig. 3(a). Baking at 900 °C causes shrinkage of only (6.1 ± 0.8) %, which compares favorably to the shrinkages between 16 % and 56 % reported for other 3D printing methods[7,17,21]. The low shrinkage of the structures printed from HSQ is attributed to the lack of an organic component in the HSQ. The low shrinkage decreases the risk of distorting 3D-printed structures during baking.

For optical applications, the transparency of the printed silica-glass structures is one of their most important properties. To demonstrate the transparency of the 3D-printed silica glass, we printed a structure, which consisted of a ring on a substrate surface and a suspended plate above the ring. We used an optical microscope to image the ring through the suspended plate, both directly after the development and after baking at 900 °C and 1200 °C (Fig. 3(b-e)). The plate was transparent in all the cases. Baking at 1200 °C caused smoothening of the printed features, which improved the optical quality of the suspended plate and resulted in sharper optical-microscope images of the ring Fig. 3(d)). We were able to control the level of smoothening by further baking at 1200 °C (Fig. 3(e)). The glass-transition temperature of fused silica is 1200 °C[38], which is consistent with the changes observed at this temperature. Even though baking at 1200 °C is unnecessary for obtaining pure silica glass, it can be used to obtain smooth surfaces, albeit some control of the structure shape can be lost. A well-controlled smoothening can be used to improve the optical quality of the printed structures. Instead of baking, local heating of the 3D structures using a laser beam[8,11] could be used to allow heat treatments even when using substrates that do not tolerate the high baking temperatures.

The transparency of the printed silica glass makes it suitable for use in 3D optical microdevices. Such structures have been reported earlier using polymers[39], but polymers are not comparable to silica glass in stability or optical transmission. To demonstrate the 3D printing of functional microdevices with our approach, we printed an integrated optical microtoroid resonator together with a bus waveguide (Fig. 4(a)). The 3D-printing process allowed us to print the ends of the waveguide so that they were bent upwards from the substrate plane. These bends enable simple coupling of light from and to optical fibers. The 3D-printed device included about 3 µm tall pillars to hold the resonator and bus waveguide above the substrate surface to minimize optical coupling to the substrate. The resonator was characterized by measuring its transmission spectrum in the optical telecommunication C-band. The input light polarization was aligned to mainly excite the fundamental transverse magnetic ($TM_{00}$) mode in the bus waveguide (Fig. 4(b)). The measured transmission spectrum showed a clear set of resonances with the expected free spectral range (FSR) for a silica-glass microtoroid of this size, thus confirming waveguiding and coupling to the resonator (Fig. 4(c)). Depending on the coupling conditions, a varying degree of secondary resonance peaks was also observed, originating from higher-order guided modes supported by the waveguide. A similar resonance spectrum was also measured after baking the device at 900 °C (Fig. 4(c)). A slight increase in the FSR after baking may be attributed to the reduced resonator radius and material densification originating from the ~6 % shrinkage of the material during baking. This confirms the possibility to use baking at 900 °C to transform the printed material to pure silica glass without compromising device functionality.

In this work, we have, for the first time, demonstrated 3D printing of submicrometric silica-glass structures. The printed structures are transparent directly after printing, and the as-printed material quality is sufficiently high for micro-optic applications, as demonstrated by 3D printing and characterizing an integrated optical resonator. If required by the application, the printed material can be made to perfectly match commercial silica glass by baking at 900 °C. This bake causes only modest shrinkage without significantly distorting the shape of 3D-printed structures. Further baking at 1200 °C allows smoothening of the printed features. The long-term stability of silica glass is highly valued for optical applications, and its chemical and thermal stability would make it possible to deposit metals or other materials on the 3D-printed structures for further functionalization, without the risk of distorting the printed structures. Future developments could also include mixing optically[40] or magnetically[41] active materials to HSQ before printing, printing 3D silica-glass structures at the tip of an optical fiber[42], and integrating 3D-printed silica-glass structures with other microstructures produced using traditional lithography techniques. The availability of 3D-printed silica glass with submicrometric features has the potential to transform micro-manufacturing based on silica glass, which is currently limited by two-dimensional lithography.

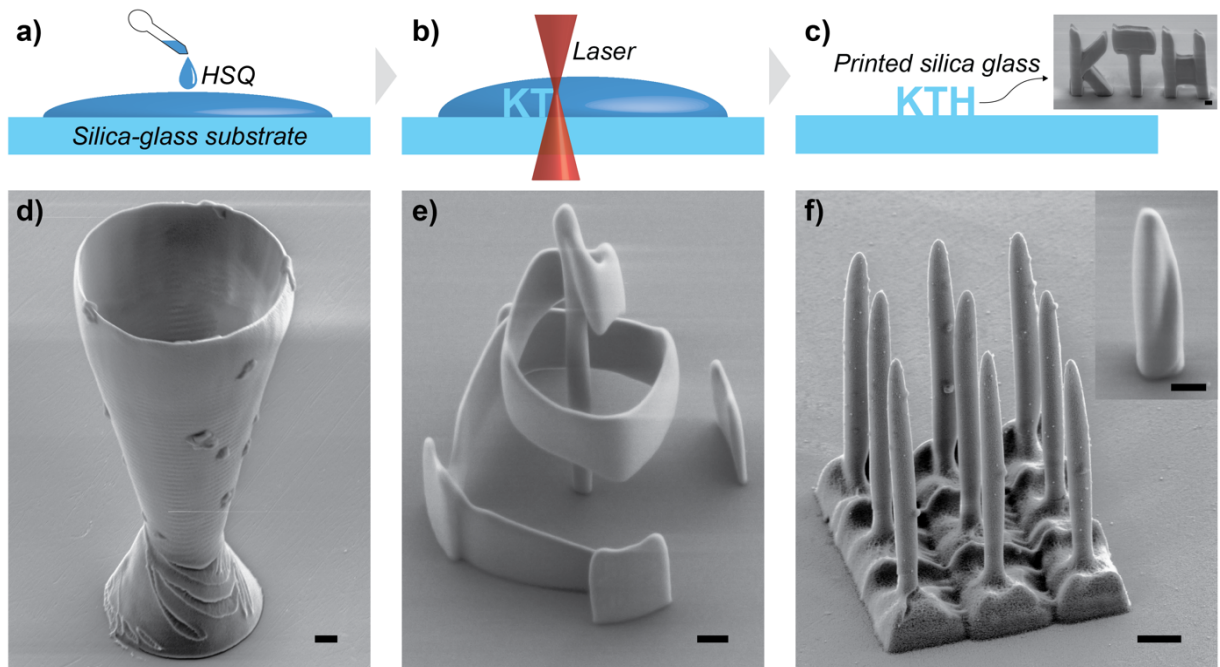

Fig. 1: The 3D-printing process together with demonstrations of the small features and complex shapes the process enables. (a) HSQ is drop-cast on a silica-glass substrate. (b) The non-linear absorption of laser light selectively cross-links the HSQ only at the focal volume. (c) Unexposed HSQ is dissolved, leaving the 3D-printed structure on the substrate. An example of a printed structure is shown in the inset SEM image. (d-e) SEM images of silica-glass structures with 3D features. (f) An SEM image of an array of 3D-printed pillars. The inset demonstrates the possibility of creating variations to the simple pillar shape. None of the printed structures in SEM images have been baked after development. The minimum lateral width of the printed structures is below one micrometer. The scale bars show a length of one micrometer.

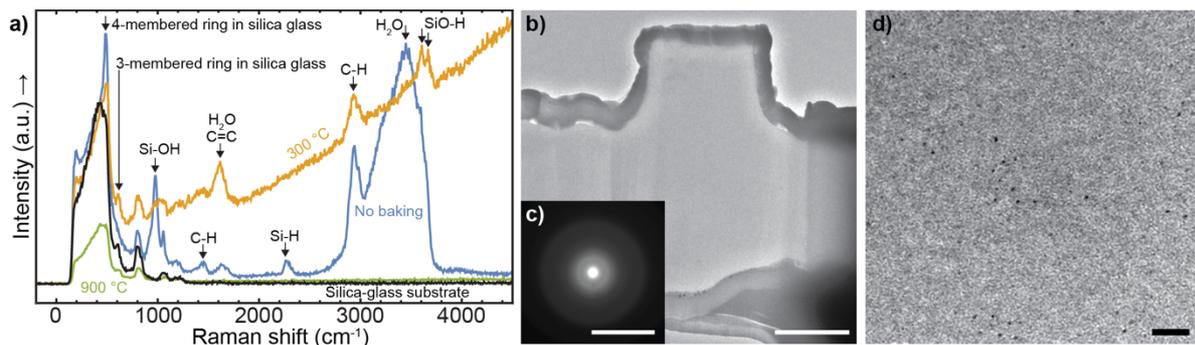

Fig. 2: Characterization of the printed material. (a) The Raman spectrum of a commercial fused silica substrate, and the spectra of the 3D-printed structures without baking and after baking at different temperatures. Raman peak positions are also summarized in Extended data table 1. (b) A scanning TEM image of a lamella sliced from the as-printed structure. No inhomogeneities or pores are observed. The darker layers surrounding the lamella are gold and platinum deposited during sample preparation. The scale bar shows 1 µm. (c) The electron diffraction pattern of concentric rings shows that the printed material is amorphous. The scale bar shows 10 nm$^{-1}$ in reciprocal space. (d) A high-resolution TEM image of the same lamella shows a low concentration of inhomogeneities, with a size of a few nanometers, visible as dark dots. The scale bar shows 10 nm.

4/13

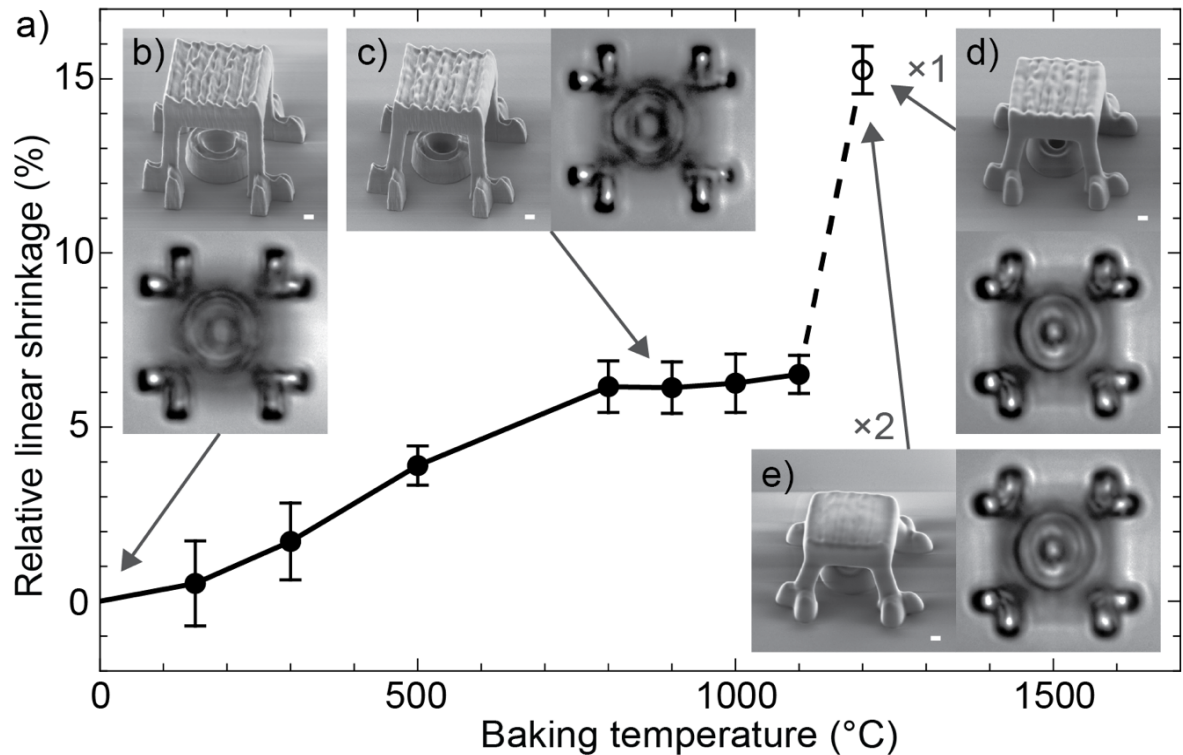

*Fig. 3: The effects of baking on structure shape. (a) The relative linear shrinkage of 3D-printed structures shown in Extended data fig. 6 after baking at different temperatures. The error limits indicate the largest measured variation from the mean value or the measurement uncertainty, whichever was larger. After baking at 1200 °C, the softening of the printed material has rounded the shape of the structures, which makes the length change incomparable to the rest of the measurements. (b) An SEM perspective view of a 3D-printed plate suspended on four legs with a ring structure underneath. Below the SEM image is an optical-microscope top-view of the ring-shaped structure imaged through the transparent suspended plate. (c) The same structure after baking at 900 °C for one hour. (d) The same structure after additional baking at 1200 °C for one hour. (e) Baking one more hour at 1200 °C causes further smoothening and rounding the structure shape. All the scale bars in the SEM images show 1 μm.*

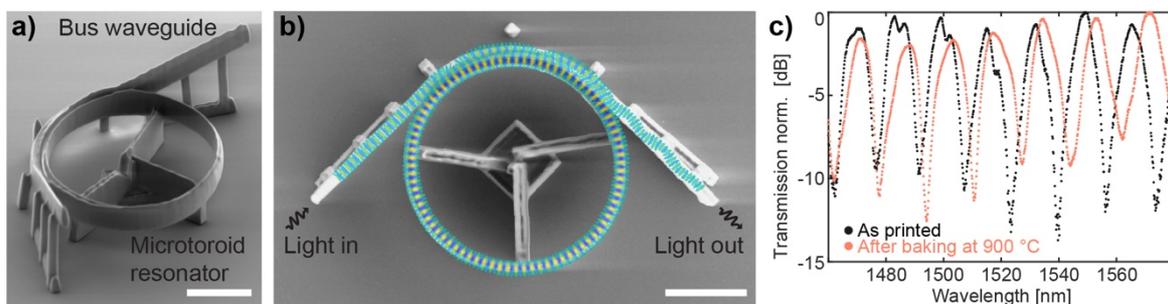

*Fig. 4: A 3D-printed microtoroid resonator. (a) An SEM perspective view of a 3D-printed photonic microtoroid resonator with a bus waveguide. The SEM image is taken from an as-printed device (i.e., before baking). The scale bar shows 10 μm. (b) An SEM top view of the same device. The SEM image is superimposed with the simulated vertical component of the electrical field at resonance when injecting the fundamental transverse magnetic ($TM_{00}$) mode into the bus waveguide. The light in the bus waveguide consequently couples to the microtoroid resonator. The scale bar shows 10 μm. (c) Measured transmission spectra through the device in the C-band, as printed, and after baking the structure at 900 °C. Both measurements were referenced to the spectrum maximum. The resonances confirm coupling into the microtoroid resonator. The free spectral range (FSR) of resonances increases after baking.*

# Methods

Preparing an HSQ layer on a substrate
A fused-silica substrate (JGS2 optical-grade fused quartz, MicroChemicals) with a thickness of 250 μm was used in the experiments. The manufacturer reports the substrate to be transmissive between 270 nm and 2 μm and to have a typical hydroxyl (OH) concentration below 300 parts per million[43]. The substrate was cleaned by rinsing first with acetone followed by isopropanol. HSQ in methyl-isobutyl-ketone-based solution (FOX16, Dow Corning) was drop-cast on the substrate. The thickness of the HSQ layer was grown to an order of 100 μm by drop-casting multiple times on the same location while allowing a few minutes for drying between the casts. After drop-casting, the sample was left to dry in a fume hood overnight. After drying, the HSQ layer was hard, and cracks appeared in the layer.

Laser direct writing of 3D-printed structures
Dry HSQ was exposed using a sub-picosecond laser (Spirit 1040-4-SHG, Spectra-Physics of Newport Corporation) operating at 1040 nm central wavelength, 10 kHz repetition rate, and producing pulses with a duration of 298 fs. The laser was focused on the HSQ through the substrate using an objective (Olympus Plan Achromat RMS40X) with a numerical aperture of 0.65. Suitable laser power was found by observing the appearance of the patterned structures through the objective using a camera. The single-pulse energies used in the patterning were between 14 nJ and 18 nJ and were measured after the pulses exited the final focusing objective of the laser system. The sample was moved by a 3-axis linear motorized stage (XMS100, Newport) and the movement speed during printing was typically between 0.5 μm/s and 1 μm/s.

Development of the 3D-printed structures
The HSQ not exposed to laser light was removed in a development step. The development was done by immersing the sample in a 0.1M solution of potassium hydroxide (Sigma-Aldrich) in de-ionized water. To this mixture, 0.05 vol% of Triton X-100 (LabChem Inc.) was added as a surfactant to decrease the size of bubbles formed in the development process to reduce the damage caused by bubbles to the printed structures. The development was continued overnight (at least 8 hours) and the sample was rinsed with de-ionized water. Most of the samples were left to dry naturally, but the sample in Fig. 1(f) was dried using critical-point drying to avoid bending or breaking the structures by surface tension.

Baking
Baking at different temperatures was done in an oven with an air atmosphere. Samples were placed in the room-temperature oven after which the oven was heated to the baking temperature. The reported temperatures are the measured air temperature inside the oven. The heating of the oven from room temperature to 1200 °C took about one hour and 51 minutes. The oven was kept at the desired baking temperature for one hour after which the oven was powered off and left to cool down naturally for about four hours. The samples were not removed from the oven before the temperature had sunk below 150 °C.

Raman and photoluminescence
Raman and photoluminescence experiments were conducted using a confocal Raman microscope (WITec alpha 300R) equipped with a 404 nm laser coupled to the microscope using a single-mode optical fiber. The excitation power was manually set below 5 mW preventing any thermal damage to the samples while a reasonable measurement time was maintained. The collected light was guided to a fiber-coupled 300 mm ultrahigh-throughput spectrometer (UHTS 300). A 600 g/mm grating was used to disperse the collected light onto a CCD camera. This setup provided an energy resolution of about 3 $cm^{-1}$, suitable for the measurement of broad photoluminescence signals. For a high lateral resolution, a 100X objective with a numerical aperture (NA) of 0.9 was used. With this objective, the signal acquisition from the focal planes above and below the focus point of the objective can be suppressed, thus confining the maximum signal intensity to a vertical space with a height of about 0.5 μm. This vertical resolution allows linking the recorded signal to a certain measurement height in the sample. The 3D-printed structures characterized were blocks suspended around 7 μm to 9 μm above the substrate. A vertical line scan starting at the substrate surface and moving the focal plane upwards through the 3D glass structure was then used to distinguish the Raman and photoluminescence signals from the substrate from those originating from the 3D-printed structures.

Length change during baking
Five T-shaped structures were used for characterizing the shrinkage (Extended data fig. 6). The lengths of the horizontal beams of the T-shaped structures were measured using SEM before baking and after baking at each temperature. The distances between the vertical pillars of the T-shaped structures were used for scaling all the SEM measurements to compensate for focusing and sample orientation differences between the measurements.

The mean value of the relative shrinkage from the five test structures is shown in Fig. 3(a). The error limits shown in that figure are either the largest measured variation from the mean at each temperature or the propagated error originating from scaling and measurement accuracy uncertainties, whichever was larger. The relative length changes of the individual test structures are shown in Extended data fig. 6.

Microtoroid resonator design
The resonator was designed using both 2D Finite Difference Eigenmode waveguide simulations (Lumerical MODE 2020a), and 3D Finite Difference Time Domain device simulations (Lumerical FDTD 2020a). The dimensions of the waveguide's cross-section were chosen to be 1.2 μm x 2.5 μm, thus supporting three transverse electric (TE) and three transverse magnetic (TM) modes in the optical communications C-band (from 1530 nm to 1565 nm). Smaller dimensions would have increased losses caused by scattering due to the supports of the bus waveguide and the microtoroid resonator. The radius of the toroid was chosen to be 15 μm as a suitable trade-off between writing time and free spectral range (FSR). A short FSR is desirable to confirm coupling into the microtoroid resonator because it allows capturing multiple resonances within the wavelength tuning range of our laser. The FSR was computed using the equation

$$FSR = \frac{\lambda^2}{n_g 2\pi r},$$

where $\lambda$ is the wavelength, $n_g$ is the group index of the guided mode, and $r$ is the radius of the toroid. The group indexes of the fundamental TE and TM modes were extracted from simulations. With a wavelength of 1550 nm, we expected an FSR of 16 nm for the fundamental TM mode, which results in six or seven observable resonances. The gap between the 3D-printed waveguide and the toroid was designed to be 500 nm and the length of the coupling region was designed to be 23 μm. These dimensions ensured the coupling of all the modes between the waveguide and the toroid, but the dimensions were not selected to achieve critical coupling. Finally, the tips of the bus waveguide were directed orthogonal to each other to prevent light from the input fiber to be reflected directly to the output fiber by the substrate surface (Fig. 4(a-b)).

Resonator optical characterization
Light for the transmission experiments was produced by a tunable laser source (Agilent 8164A) with fixed power of 1 mW and operating at wavelengths between 1457 nm and 1582 nm. The transmitted light was measured by a wavelength domain component analyzer (Agilent 86082A). Single-mode optical fibers were used to transfer light from the laser source to the resonator and back to the analyzer. To couple light between the fibers and the ends of the bus waveguide, we used fibers with tapered tips (AMS technologies, TSMJ-X-1550-9/125-0.25-7-5-26-2.1). The fiber tips had a working distance of 26 μm and a spot diameter of 5 μm. The fiber tips were positioned using six linear piezo stages (Newport, AG-LS25), which had a minimum incremental motion of 50 nm. Two microscopes imaging the positioning from above and from a side facilitated the precise alignment of the fiber tips to the ends of the bus waveguide. We controlled the direction of the linear polarization of the light coupled into the bus waveguide using a fiber polarization controller (Thorlabs, FPC030). This allowed us to preferably excite either TM or TE modes in the bus waveguide. The light transmitted through the resonator device was recorded in two components with orthogonal polarizations, which were separated using a beam splitter (Thorlabs, PBC1550SM). Because the bus waveguide supports multiple modes with smaller mode diameter than the fiber, the modes excited in the waveguide were sensitive to the exact positioning of the fiber tips. Excitation of higher-order modes produced extra sets of resonances in the transmission spectrum. When measuring the device transmission, we recorded spectra for multiple positions of the input fiber. The orthogonal polarization components of the spectra were recombined for analysis, and from this set of spectra, we automatically selected the one with the best fit to a single set of resonances, which suggests the strongest coupling to the fundamental mode.

## Acknowledgments

We thank Cecilia Aronsson for substrate dicing and general assistance in the cleanroom and Dr. Fredrik Gustavsson for TEM characterization. This work has been funded by Swedish Foundation for Strategic Research (SSF GMT14-0071) and by European Commission under the project Graphene Flagship 785219, 881603, and FET-Open Queformal 829035.

# Extended Data

*Extended data table 1: Raman shifts of the observed Raman peaks*

| Peak [reference] | Observed Raman shift ($cm^{-1}$) |
|---|---|
| 4-membered ring [33] | 490 |
| 3-membered ring [33] | 605 |
| Si-OH [30] | 975 |
| C-H (asymmetric bending) [44] | 1455 |
| C=C [44] | 1620 |
| $H_2O$ (bending mode) [30] | 1620 |
| Si-H [45] | 2260 |
| C-H (stretching) [31] | 2935 |
| $H_2O$ (O-H stretching) [30] | 3450 |
| Unspecified SiOH species [30] | 3598 |
| SiO-H stretching [30] | 3665 |

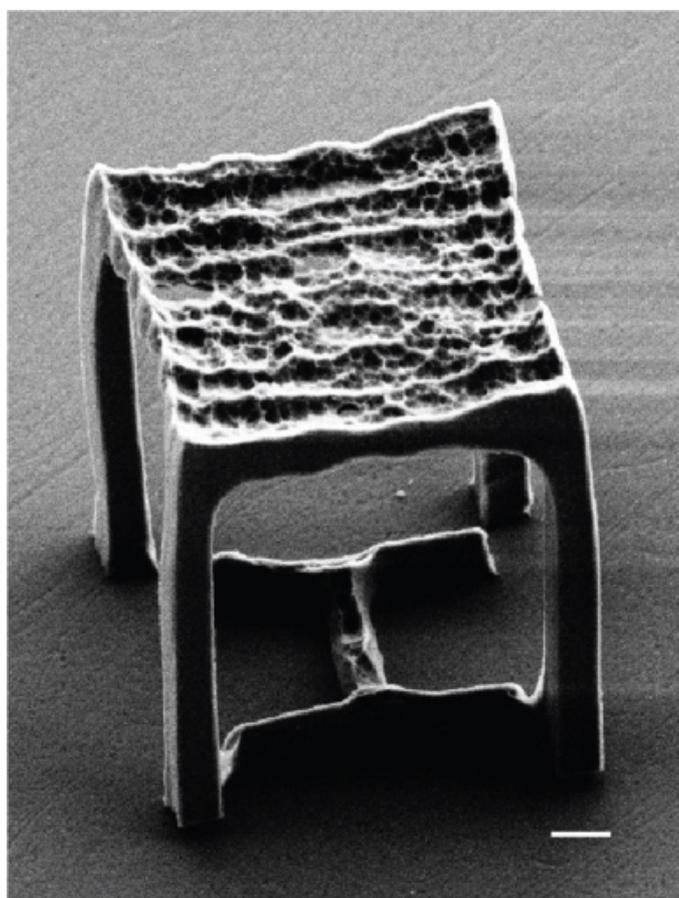

*Extended data fig. 1: A 3D-printed structure with a horizontal top layer that has a thickness of less than 1 μm. The scale bar shows the 1 μm distance. The thin structure was produced by reducing single-pulse energy.*

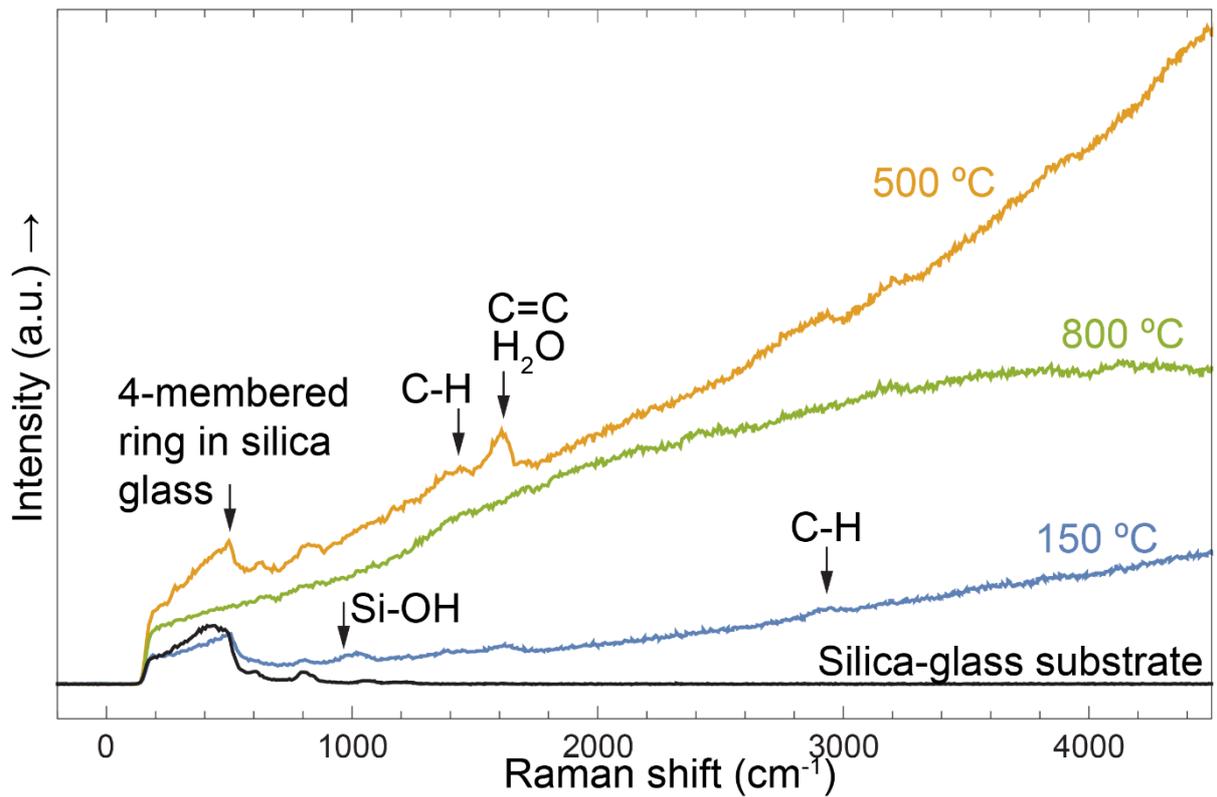

*Extended data fig. 2: Raman spectra measured from a silica-glass substrate and 3D-printed structures after baking at the temperatures not presented in Fig. 2(a). The strong photoluminescent background makes it difficult to identify Raman peaks from the spectra.*

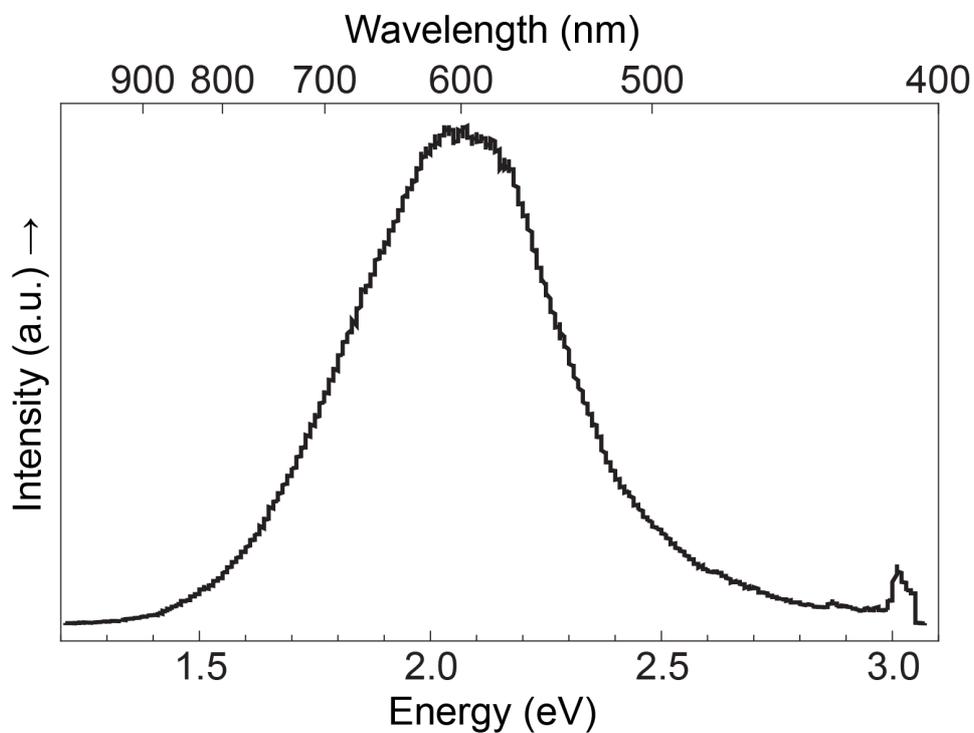

*Extended data fig. 3: A photoluminescence spectrum measured from the sample baked at 500 ºC. The peaks around 3 eV originate from Raman scattering.*

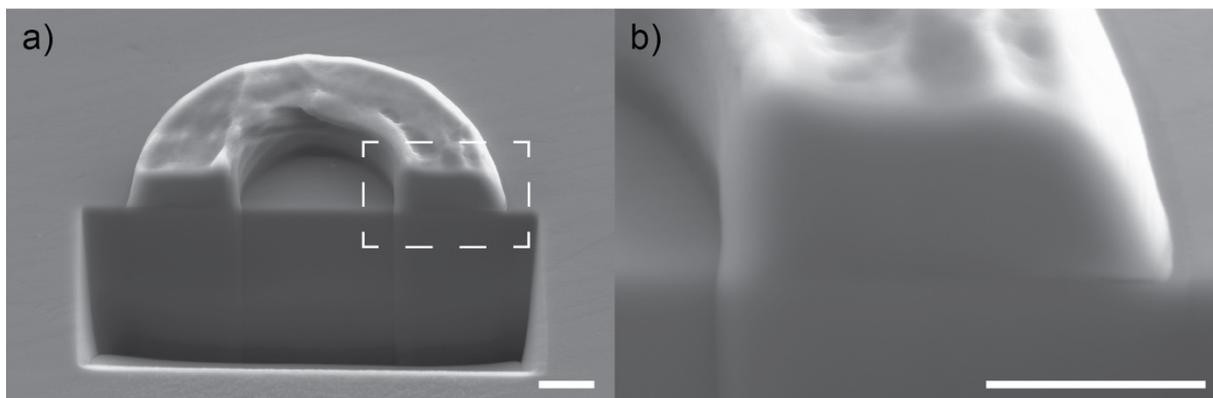

*Extended data fig. 4: (a) An SEM image of a laser-printed structure cross-sectioned by focused ion beam milling. (b) An enlarged view of the cross-section. No porosity is visible in the cross-section. Scale bars show 1 µm.*

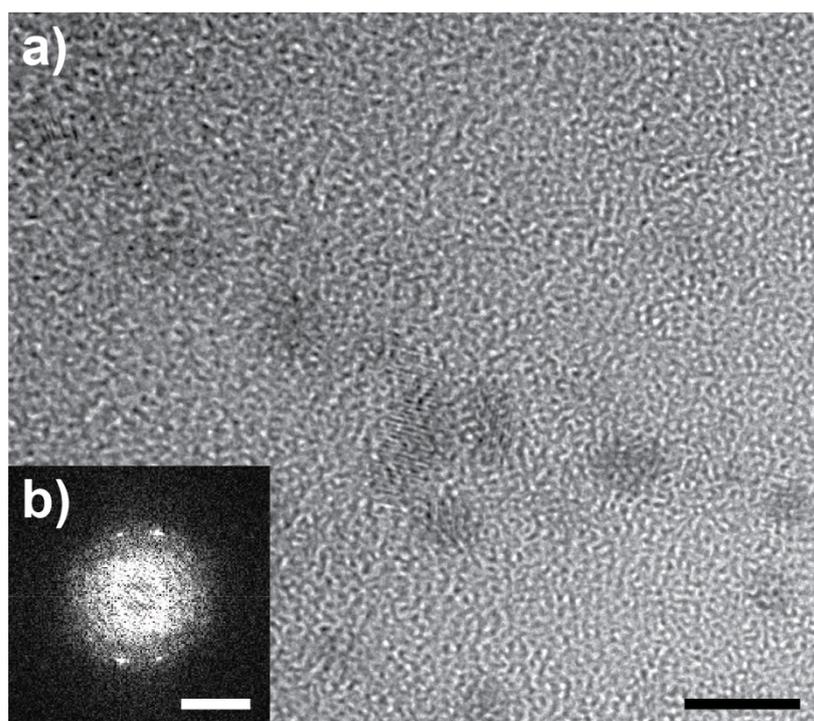

*Extended data fig. 5: (a) A high-resolution TEM image of as-printed material. The dark inhomogeneities show a fringe pattern indicative of crystalline material. The scale bar shows 5 nm. (b) An electron diffraction pattern from an inhomogeneity shows a diffraction pattern of crystalline material. The scale bar shows 5 $nm^{-1}$ in reciprocal space.*

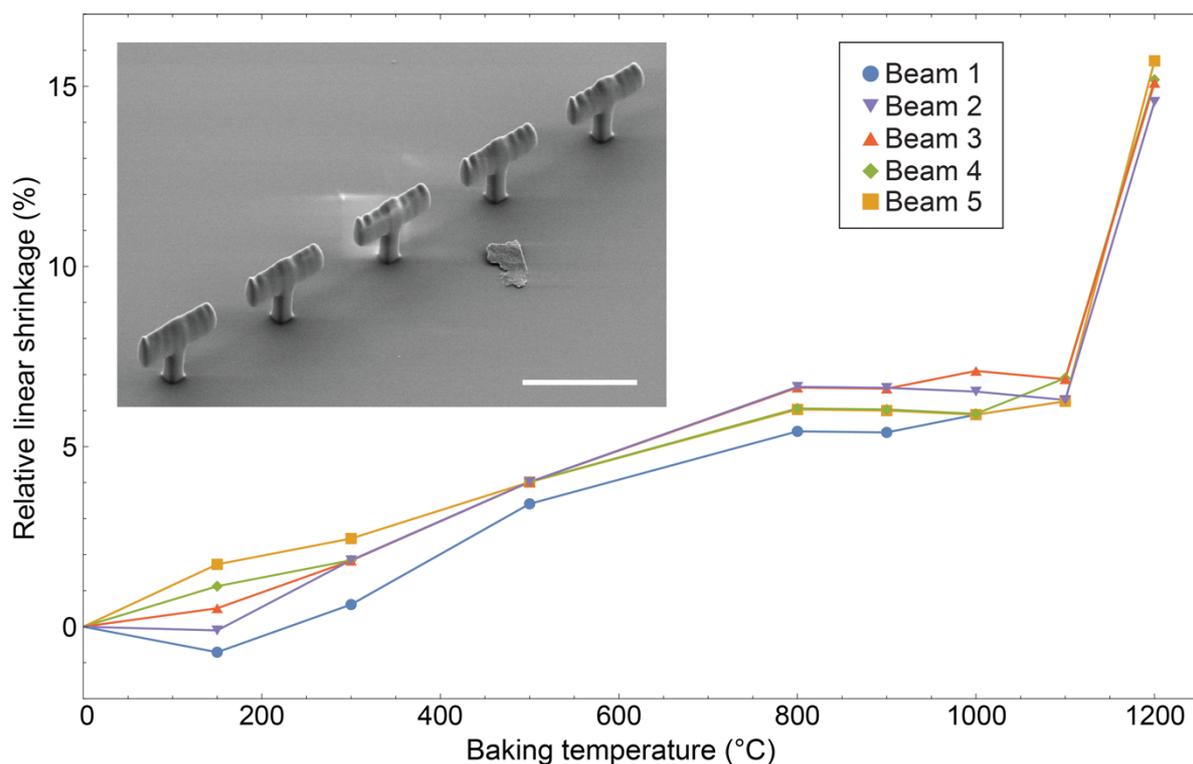

*Extended data fig. 6: The relative linear shrinkage of each separate test beam after baking at different temperatures in comparison to their lengths before baking. The inset shows an SEM image of the five test beams after baking at 1200 °C. The scale bar shows 10 μm.*